\documentclass[preprint,aps]{revtex4}
\usepackage{graphicx}
\usepackage{booktabs}
\usepackage[usenames, dvipsnames]{color}
\usepackage{multirow}
\usepackage{setspace}
\usepackage{amsmath, amsthm, amssymb}
\usepackage{rotating}
\usepackage[ pdftex, plainpages = false, pdfpagelabels, 
                 pdfpagelayout = useoutlines,
                 bookmarks,
                 bookmarksopen = true,
                 bookmarksnumbered = true,
                 breaklinks = true,
                 linktocpage=all,
                 pagebackref=false,
                 colorlinks = true,
                 linkcolor = BrickRed,
                 urlcolor  = blue,
                 citecolor = BrickRed,
                 anchorcolor = green,
                 hyperindex = true,
                 hyperfigures
                 ]{hyperref} 
                 
\newcommand{\comment}[1]{}

\newcounter{figref}

\begin{document}

\title{\href{http://necsi.edu/research/social/foodprices/update/}{UPDATE February 2012 --- The Food Crises: Predictive validation of a quantitative model of food prices including speculators and ethanol conversion}} 
\date{\today}  
\author{Marco Lagi, Yavni Bar-Yam, Karla Z. Bertrand and \href{http://necsi.edu/faculty/bar-yam.html}{Yaneer Bar-Yam}}
\affiliation{\href{http://www.necsi.edu}{New England Complex Systems Institute} \\ 
238 Main St. Suite 319 Cambridge MA 02142, USA \vspace{2ex}}

\begin{abstract}
Increases in global food prices have led to widespread hunger and social unrest---and an imperative to understand their causes. 
In a previous paper published in September 2011, we constructed for the first time a dynamic model that quantitatively agreed with food prices. Specifically, the model fit the FAO Food Price Index time series from January 2004 to March 2011, inclusive. The results showed that the dominant causes of price increases during this period were investor speculation and ethanol conversion. The model included investor trend following as well as shifting between commodities, equities and bonds to take advantage of increased expected returns. Here, we extend the food prices model to January 2012, without modifying the model but simply continuing its dynamics. The agreement is still precise, validating both the descriptive and predictive abilities of the analysis. 
Policy actions are needed to avoid a third speculative bubble that would cause prices to rise above recent peaks by the end of 2012. 
\end{abstract}

\maketitle

Recent increases in basic food prices are severely impacting vulnerable populations worldwide. Proposed causes such as shortages of grain due to adverse weather, increasing meat consumption in China and India, conversion of corn to ethanol in the US, and investor speculation on commodity markets lead to widely differing implications for policy. A lack of clarity about which factors are responsible reinforces policy inaction.

In a paper published in September 2011, we constructed for the first time a dynamic model that quantitatively agreed with food prices \cite{food_prices}. The model was able to fit the FAO Food Price Index time series from January 2004 to March 2011, inclusive, and showed that the dominant causes of price increases during this period were investor speculation and ethanol conversion. It included investor trend following as well as shifting between commodities, equities and bonds to take advantage of increased expected returns. 

\begin{figure}[tb]
\refstepcounter{figref}\label{fig:food_fit}
\href{http://necsi.edu/research/social/foodprices/update/food_spec_2_2012.pdf}{\includegraphics[width=0.9\linewidth]{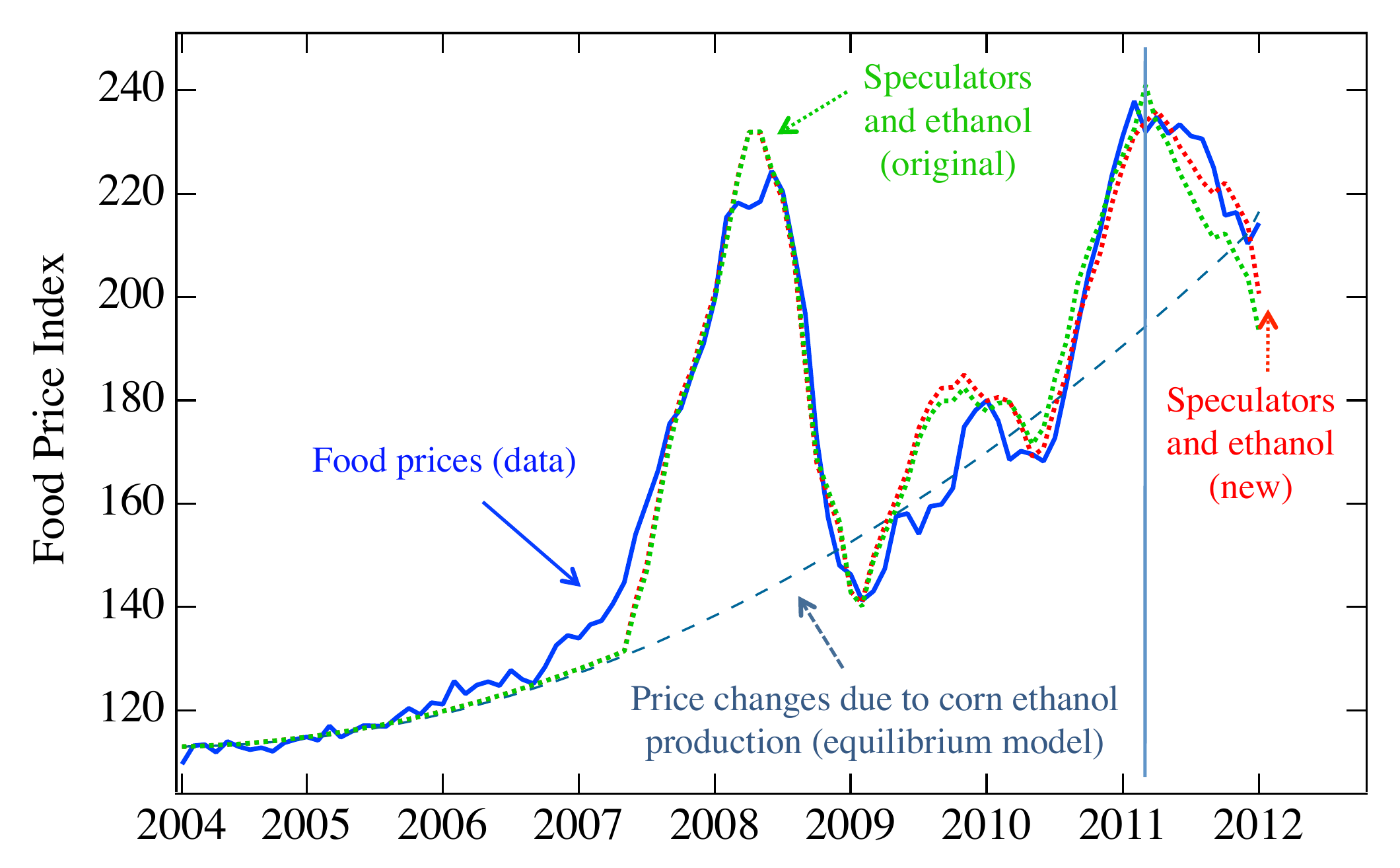}}
\caption{\textbf{Food prices and model simulations} - The FAO Food Price Index (blue solid line) \cite{source_fao}, the ethanol supply and demand model (blue dashed line), where dominant supply shocks are due to the conversion of corn to ethanol so that price changes are proportional to ethanol production (see \cite{food_prices}, Appendix C) and the results of the speculator and ethanol model (green and red dotted lines), that adds speculator trend following and switching among investment markets, including commodities, equities and bonds (see \cite{food_prices}, Appendices D and E). The green curve is the fit extended to the present with the original parameter values, the red curve is the fit with new optimized parameters. The vertical blue bar marks the end of the original fit in March 2011. Original parameters \cite{food_prices}: $k_{sd} = 0.098$, $k_{sp} = 1.29$, $\mu_{equity}\gamma_0 = - 0.095$, $\mu_{bonds}\gamma_0 = -67.9$. New optimized parameters: $k_{sd} = 0.093$, $k_{sp} = 1.27$, $\mu_{equity}\gamma_0 = - 0.085$, $\mu_{bonds}\gamma_0 = -48.2$.}
\end{figure}

Here we show that the model is still able to fit food prices up to currently available data, i.e. January 2012. In Fig. \ref{fig:food_fit} we plot both the original fit (green curve), extended to the present using the original parameters, and a new fit obtained by adjusting the parameters to optimize the fit over the entire period (red curve). Without any modification to its assumptions or formulation, by extending its original dynamics, the model proves to be robust, and consistent with the ongoing behavior of food prices. The agreement of the fit with the FAO Food Price Index is still strikingly quantitatively accurate, validating both the descriptive and predictive abilities of the model.

We note that the direction of food price changes at the time of the previous report has reversed since then in both the data and the model. The values of the new fitting parameters did not change substantially with respect to the old ones (see Fig. \ref{fig:food_fit} caption). Parameters with large impacts on prices, the supply and demand equilibrium restoring (Walrasian) parameter $k_{sd}$ and the speculator parameter $k_{sp}$, changed by $5\%$ and $2\%$ respectively. The smaller impact parameter that controls switching of investment to equities $\mu_{equity}\gamma_0$ changed by less than $10\%$. Finally, the change of 30\% in the parameter that controls investment switching to bonds has almost no impact because it multiplies small changes in US 10-year Treasury Note Yield. The consistency of the results implies that there have been no major changes in either the supply and demand or the trend following aspects of the model. However, it is worth understanding what changes in the underlying assumptions may give rise to deviations between data and model predictions in the future. 

At the end of 2011 ethanol subsidies expired \cite{Pear2012}, the result of legislative actions over the previous year \cite{Krauss2011}. Even though recent projections show that ethanol conversion might not increase as rapidly as before \cite{source_grains}, persisting high oil prices could continue to drive ethanol conversion, and a government-guaranteed demand for 37\% of the US corn crop is still in place \cite{Smith2012}. The question of whether the conversion of corn to ethanol will continue to rise is critical. If the current trend continues, later this year or early next year grain prices will reach levels sufficient to cause widespread social unrest \cite{food_crises}, even without the bubbles caused by speculation. 

The impact of trend following speculation may change in the future if the size of the investor pool in the commodity markets changes. The quantitative accuracy of the model prices implies that the pool has been essentially constant from 2007 until the end of 2011. New policies on speculative activity by the US Commodities Futures Trading Commission are scheduled to come into effect by the end of 2012 \cite{FedReg}. Their impact will depend on details of the rule that is adopted---the specific position limits and how they are applied across different commodity based futures, options and swaps.

The ability of our model to characterize the non-equilibrium behavior of markets is an important extension of economic analyses that have previously not been designed for this purpose (see Appendix A in \cite{food_prices}). The current trend of prices suggests that in the immediate future market prices may become lower than equilibrium, consistent with bubble and crash market oscillations. This may provide a short term reprieve from the extremely high food prices that have caused a global food crisis \cite{FAOreport} and related social unrest---the timing of food riots and the revolutions in North Africa and the Middle East coincides with the recent food price peaks \cite{food_crises}. Still, the current equilibrium value is about 50\% higher than the prices prior to the impact of the ethanol shock. If the ethanol production continues to grow according to its multiyear trends, even the equilibrium price will reach social crisis levels in just one year \cite{food_crises}. The recent expiration of ethanol subsidies in December of 2011 may not be adequate to prevent this outcome. Moreover, whether the ethanol conversion continues to increase or not, based upon the value of the speculator parameter another speculative bubble is expected within a year. By the beginning of 2013 this bubble would increase prices above those of the price peaks in 2008 and 2011. Policy actions are possible to limit commodity speculation and ensure ethanol production returns to levels that do not severely impact food prices. 

This work was supported in part by the Army Research Office under grant \#W911NF-10-1-0523.

\end{document}